# Developing an Index of National Research Capacity

Caroline S. Wagner[1]* and Travis A. Whetsell[2]

[1]John Glenn College of Public Affairs, The Ohio State University, Columbus, Ohio, USA. Wagner.911@osu.edu. ORCID: 0000-0002-1724-8489; *corresponding author
[2] School of Public Policy, Georgia Institute of Technology, Atlanta, Georgia, USA. travis.whetsell@gatech.edu, ORCID: 0000-0001-5395-4754

**Abstract**

We test the feasibility of adding broad social, political, and governance indicators to national metrics as a way to enrich assessment of national research capacity. We factor analyze two sets of variables for 172 countries from 2013 to 2021, one being traditional measures associated with national science and technology capacity such spending, and a second being broader social, political and governance measures, such as academic freedom. As expected, two factors emerge, one for raw or "core" research capacity and the other indicating the wider governance context. Further analysis shows convergent validity within the two factors and divergent validity between them. Nations rank differently when governance context is included. Ranks also vary as a function of the chosen aggregation method. As a test of the predictive validity of the capacity index, we find both factors to be associated with country level field weighted citation index. Policymakers and analysts may find stronger feedback from this approach to quantifying national research strength.

1. **Introduction**

This paper combines traditional research and development (R&D) measures with data representing the larger context of governance to provide richer insight into research capacity of nations. This research is a preliminary attempt at exploratory data analysis guided by a framework that extends research and development (R&D) measures by adding governance indicators to enrich assessment of national research capacity. The effort is intended to aid researchers in comparing nations on policy-related features rather than the business-oriented measures common among other indices.

The scholarly literature catalogues measures of research capacity focusing primarily on research and development spending, public investments in infrastructure, and institutions. Lundvall (2007) characterizes the dynamic interaction of factors, including those from political, economic, and social contexts, as "national systems of innovation" (NSI) . In a similar approach to Furman et al. (2002) who distinguish between core capacity and the environment for innovation, Lundvall (2007) defines inputs to national systems as having "core" and "wider context" elements. We broaden the purview of traditional research capacity indexes by including attention to these governance factors. Practitioners may find this approach useful when comparing national science and technology policies, while scholars may find it useful to model and analyze elements of the global research system.

This article builds upon previous attempts to assess national capacity for knowledge creation, variously termed 'impact measures', 'wealth creation', or 'research productivity'. The proposed methodology adds to, and aggregates, data used in earlier efforts, where indicators have not been indexed. We accept existing indicators as given and then gather them into a new index.

We undertake this work for three reasons: 1) existing indices do not provide useful feedback to the national, public-sector policymakers; 2) data have become more broadly available and provide an opportunity to include more countries; 3) methods are needed to better compare nations on their capacity to conduct research. The application of composite indices in policy analysis has been widely accepted as a pragmatic approach for evaluating national performance.This acceptance is attributed to the ability of indices to "effectively encapsulate intricate and occasionally evasive matters across various domains" (Nardo & Saisana, 2009, p.2).

Following this introduction, the paper is structured in four parts. The literature review justifies the choice of metrics and indicators for this paper. Then, the methodology for constructing the index, the data chosen and applied, and the statistical analyses utilized are described. Following the discussion of the methodology and data, we present the results of statistical analyses along with a discussion of the limitations. In the conclusion, the results are discussed focusing on the index's applicability to future research, and the next steps for using the index to compare nations.

## 2. Literature Review

First, we compare existing innovation indexes and explain why our proposed index fills a gap. Then, we draw from literature to present a conceptual framework that categorizes our variables into three broad categories: 1) raw scores of research capacity, Lundvall's "core capacity"; 2) broad measures of social and political context, Lundvall's "broader context"; and 3) outcomes of science, specifically, citation impact. In contrast to a standard literature review approach which presents testable hypotheses, our approach is to mine the literature for candidate indicators of national research capacity.

*2.a. Existing Innovation Indexes*

Indices are a construct of broad interest. Indexes of national innovation, competitiveness, and knowledge are compiled each year to provide input to business and public policy by ranking countries on baskets of variables. These reports often focus on economic growth, business, and commerce, with little effort to carve out the global from the national data. From a policy perspective, indexes such as these have limited value as feedback to R&D policy support. Of the business-oriented indexes, the most prominent are 1) the Global Competitiveness Index (GCI), 2) the Global Innovation Index (GII), and 3) the Global Knowledge Index (GKI), each of which take a slightly different approach to measuring innovation. Our index does not draw upon these reports, but we summarize them here to contrast our approach with theirs.

1) The World Bank publishes the Global Competitiveness Index as a publication of the World Economic Forum Global Competitiveness Report (GCR); this work was most recently issued in 2019[1]. The GCI draws upon 110 variables and a survey of businesspeople, covering 141 countries. GCI variables are grouped into "pillars" with nested indices covering human capital; market conditions; policy environment and enabling conditions; technology and innovation; and physical environment. The outcomes provide ranks of countries by their competitiveness.

2) The Global Innovation Index (GII) is co-published annually by Cornell University, INSEAD Business School, and the World Intellectual Property Organization (WIPO). The GII is a multilevel index with 81 variables, nested into sub-indices around seven pillars: institutions; human knowledge and research; infrastructure; market sophistication; business sophistication; knowledge and technology outputs; and creative outputs. GII covers the economies of 132 countries in the 2022 report.

---

[1] In the 2000s, the World Bank created the Knowledge Economy Index, which grew from a report that described a Knowledge Assessment Methodology (Chen & Dahlman, 2006); the World Bank no longer publishes the KEI index.

3) The United Nations endorses the publication of the Global Knowledge Index (GKI) compiled by the Mohammed Bin Rashid Al Maktoum Knowledge Foundation (2022). GKI is a multilevel index combining 199 variables, nested in sub-indices, covering seven areas: pre-university education; higher education; research, development, and innovation; information and communications technology; economy; enabling environment, consisting of governance, socio-economic, and health and environment. GKI covers 138 countries.

These reports include national data (such as exports) together with data that have a high degree of globalization (such as international patents) which does little to help science policymakers assess national policy impacts on research strengths.

Government support in the form of grants and research infrastructure provides aid to create non-rivalrous goods to enrich a "pool of knowledge" from which users can draw and which spill over to private sector users. While we cannot perfectly disentangle rivalrous from non-rivalrous knowledge, our approach limits indicators to those that feed back to public knowledge specific to the country. We do not include trade data, international patents, or global mobility. Further, we use scientometric measures that disentangle national versus collaborative capacity by fractional counting.

Li, Zhang, and Liu (2020) developed a citation-based scientific capability identification (CISCI) tool to examine national capability in what they termed "dual science roles" by examining both cited and citing behavior in a networked structure. They find, for 158 countries, different contributing roles for different fields of science; they further show that these roles and rankings change over time. The USA, Canada, and the UK were consistently highly ranked by their tool, with rapid improvements for China. This admirable approach provides useful insights into national capacities, but we have the further goal of testing the wider context of governance, which is not covered by Li et al. (2020).

### 2.b. Measures of Research and Development Capacity

This section reviews the literature on R&D indicators which are used in our index. May (1997) compared the scientific wealth of nations for high-performing countries by calculating the number of scholarly articles along with citations to these articles on a nation-by-nation comparison. May found that 15 nations accounted for 81% of all scholarly articles which has expanded since his article. King (2004) conducted an analysis like May (1997), examining the national distribution of the top 1% of most highly cited papers, finding the United States to be dominant at that time, a finding that has also changed. Wagner and Jonkers (2017) compared nations on international engagement and found a positive relationship between open exchange and the impact of and quality of science, supporting earlier work by Barro (1996).

Previous studies including Barro & Lee (1994;) Martin (1996); Frame (2005); Cole & Phelan (1999) identified and assessed the usefulness of indicators of knowledge-creating capacity, many of which are codified in the Frascati Manual (OECD, 2015). A consensus emerged around spending on research and development and the number of trained individuals as core indicators of R&D capacity (Romer, 1989). Others have added regulatory quality and political stability (Furman et al., 2002; Lundvall, 2005). Others measured knowledge production (number of published articles) (May, 1997) and the number of domestic patent registrations (Tong & Frame, 1994; Narin, 1994; Schmoch, 2004; Lepori et al., 2008).

Analysts often include the number of research-conducting academic and non-academic research institutions in discussions of national capacity (Knack & Keefer, 1995). International cooperation, fractionally counted to attribute numbers to participating countries, is also commonly used as an indicator of global engagement and openness (OECD, 2015; Wagner & Jonkers, 2017).

Research and development expenditure is a major contributor to the knowledge economy, and it is always included in any measure of research capacity. Gross expenditure on research and development (GERD) is universally recognized in the literature as a key indicator of knowledge-creating capacity, and there is a strong correlation between R&D spending and economic growth (Howitt, 2000; Salter & Martin, 2001). Gulmez and Yardimcioglu (2012) discovered positive effects of R&D spending on the income of 21 OECD member nations between 1990 and 2010, demonstrating that a one percent increase in R&D spending led to a 0.77 percent increase in economic growth. Other authors have studied the relationship between scientific investment and national development and discovered a significant correlation between R&D investment and both short-term and long-term growth in developing and developed nations (Gittleman & Wolff, 1995; Goel & Ram, 1994; Gumus & Celikay, 2015). Adams (1990), May (1997), and King (2004) demonstrate significant correlations between R&D expenditures and economic expansion.

Education and human resources are included in any assessment of national research capacity. In earlier attempts to measure national scientific activity, Frame (2005) emphasized the importance of education, while Barro (1991), citing work by Nelson and Phelps (1966), showed that nations with highly trained human capital were better able to absorb new products or ideas, which we include as 'academic institutions' and 'non-academic research institutions'. Fedderke (2005), supports Romer (1990), noting that quality, rather than quantity, of human capital contributes to total productivity at the national level. Capacity for training, education, and knowledge transfer is also essential: Schofer et al. (2000) showed that the size of scientific labor sources and training systems had a positive effect on national economic growth. Cole and Phelan (1999) identified a positive relationship between the number of research scientists and economic growth, and Barro (1991) showed that GDP growth is positively related to availability of trained human capital.

Scholarly productivity is an imperfect indicator of research capacity. The number of scholarly articles published is widely used as an indicator of the strength of a national research sector (OECD, 2015; Martin, 1996). Recent research by Miao et al. (2022) demonstrates a correlation between a nation's scientific output and its economic growth and complexity (see also Cilliers, 2005; Ahmadpoor & Jones, 2017). Past productivity forms a basis for expectations of productive capacity in the future. This finding supports the findings of Cimini et al. (2014), who discovered that OECD member states had more diverse research systems when measured in articles than developing countries (see also, OECD (2021)).

Patent counts are used as indicators for technological development or entrenched knowledge (WIPO Patentscope, 2022). Crosby (2000) identified a positive correlation between the number of patents and economic growth, while Kogan et al. (2012) support Furman et al. (2002) demonstrating that the scientific content of the patent is positively correlated with the patent's value to the economy. Patent offices often differentiate between national and international patents. National patents indicate inventive capacity, while international patents indicate the attractiveness of an economy as a market for goods. Our index uses only national patents.

### 2.c. Research on Social and Political Context

The importance of good governance and political stability for knowledge-based economic growth is well-documented (Bäck & Hadenius, 2008). Rule of law and freedom from corruption are correlated with higher economic growth and expansion of a knowledge economy (Haggard et al., 2008). Barro (1996) and Cole & Phelan (1999) showed that growth is correlated with the maintenance of the rule of law, free markets, small government consumption, and high human capital. Other studies have shown a positive relationship between political stability, technological change, and growth (Barro, 1991; Barro & Lee, 1994; Hall & Jones, 1999) and between democracy and growth (Barro, 1996). Berggren & Bjørnskov (2022) showed correlations between academic freedom and innovation. Whetsell et al. (2021) showed the relevance of democratic governance in predicting the national performance in science, and Wang et al. (2021) show similar effects on technology. Whetsell et al. (2021) showed that levels of polyarchy, measured through Varieties of Democracy Project data (Coppedge et al., 2011; Coppedge et al., 2023), is a significant correlate of field-weighted citation impact at the national level.

There is mixed evidence regarding the role of regulations, standards, and enforcement in promoting research and development and science capacity. Some economists showed that regulatory burdens can hinder innovation, competitiveness, and national trade positions (Hahn & Hird, 1991). In contrast, Porter and van der Linde (1995), discussed in Blind (2012), suggest that, while ambitious environmental regulations may be costly for national industry at the outset, regulations may help to improve international competitiveness and increase exports of environmental technologies over the longer term. Our view aligns with that of Blind (2012) that regulations and standards aid research and innovation, and we include the Coppedge et al (2011) data on regulatory quality in our index.

The enforcement of intellectual property rights (IPR) has also garnered considerable research interest. According to Blind (2012, p. 393), innovation is supported by "...institutional regulations that ensure adequate enforcement of intellectual property rights." Blind (2012) cites Koch et al. (2004), who demonstrate that IPR regulations have an advantageous effect on the R&D intensity of the former G7 nations. Greenhalgh and Rogers (2010) demonstrated that IPR enforcement serves as an indicator of the quality of research. The rule of law facilitates the invention and innovation processes in both the public and private research sectors.

The ability of researchers to access new ideas, or diffusion, as critical to research capacity and success is supported across many parts of the literature, such as Björk & Magnusson (2009), exploring the role of interactions among researchers as tied to innovation; Lopez-Vega, et al. (2016) ask where and how to search, focusing on the Internet in their article on that topic. The role of the Internet in improving and enhancing search has received attention in high-level policy documents such as OECD's report, "Economic and Social Benefits of Internet Openness" (2016). We The open Internet measure used in our index is drawn from Varieties of Democracy data.

### 3. Variables, Data, and Methods

As a practical starting point for constructing composite indexes, the OECD Handbook on Constructing Composite Indicators provides guidelines (Nardo & Saisana, 2009). The handbook suggests the following actions: 1) establish a theoretical framework; 2) choose variables; 3) impute missing data; 4) conduct multivariate analysis; 5) normalize data; 6) weight and aggregate data; and 7) present findings. This section will discuss variable selection, data sources, and analysis techniques, as the theoretical structure has already been covered.

### 3.A. Choice of Indicator Variables and Data Sources

Table 1 provides a description of all the variables used in the analysis. FWCI is used as a dependent variable in later regression models. The remaining variables listed in Table 1 identify the proposed national research capacity index: gross research and development spending (RD), raw number; number of resident patent applications (ResPatents); number of academic institutions affiliated with publications (AcadInst); number of non-academic research institutions affiliated with publications (NonAcadInst); number of unique authors listed on publications (Authors); number of publications fractionally counted by nation (Pubs); number of papers that are international collaborations fractionally counted by nation (IntlPubs); open internet access (OpenInternet); rule of law (RuleLaw); regulatory quality (RegQuality); political stability (PolitStability); non-corruption (NonCorrupt); electoral democracy (Polyarchy); and academic freedom (AcadFreedom).

Our theoretical model suggests that these variables can be broadly categorized in terms of measures of 1) raw capacity and 2) political context of research (governance). These measures comprise a large variety of indicators often used in the analysis of country-level research capacity, investment, productivity, and output. All data were gathered from the World Bank Indicators (World Bank Databank, 2022), the Varieties of Democracy Project (Coppedge, 2023), and Scopus (2022) (an Elsevier database). The first two sources are publicly available, the database materials were made available as a courtesy to the project.

*Table 1 Variable Description and Summary Statistics*

| Variable Name | Description | Data source |
|---|---|---|
| RD | Gross research and development spending: raw number | World Bank Indicators |
| ResPatent | Number of resident patent applications | World Bank Indicators |
| AcadInst | Number of academic Institutions: paper affiliation | Scopus/Elsevier |
| NonAcadInst | Number of non-academic institutions: paper affiliation | Scopus/Elsevier |
| Authors | Number of unique authors: paper affiliation | Scopus/Elsevier |
| Pubs | Number of publications: fractional count | Scopus/Elsevier |
| IntlPubs | Number of international co-pub papers: fractional count | Scopus/Elsevier |
| OpenInternet | Country approach to regulating/controlling Internet | Varieties of Democracy |
| RuleLaw | Rule of law: crime, judicial & contract effectiveness | World Bank Indicators |
| RegQuality | Regulatory quality: burden of regulation on markets | Varieties of Democracy |
| PolitStability | Political stability: probability of gov. destabilization | World Bank Indicators |
| NonCorrupt | Control of corruption: use of public power for private gain | World Bank Indicators |
| Polyarchy | Electoral Democracy Index | Varieties of Democracy |
| AcadFreedom | Academic Freedom Index | Varieties of Democracy |
| FWCI | Fractional Field Weighted Citation Index | Scopus/Elsevier |

### 3.b. Missing Data

Finding a balance between coverage (such as the number of countries, regions, or other units) and comprehensiveness (such as the number of issues and aspects of science) is an inherent difficulty in the construction of an index. No index can be expected to include every variable on every country related to a subject. A requirement for more detailed data will necessarily lead to fewer nations, regions, or groups being included in the analysis. For example, developing countries typically collect less data than developed countries, and the available statistical data may be less reliable. A stipulation for complete data would exclude some nations.

We sought to construct our index to include as many nations as possible. However, numerous variables of interest had low data coverage. This presents an interesting problem for research. High-quality measures are generally available for countries whose status is already well known and whose research systems are well developed, while data is lacking on those that are likely to experience the greatest change over time and whose status is of particular interest.

Data from Scopus/Elsevier is comprehensive across almost all countries, so these data formed the base sample for the subsequent merging of data. The Varieties of Democracy (Coppedge, 2023) data also covered many countries, as did many World Bank indicators. However, we dropped tertiary enrolment because of low data coverage. Gross Research and Development Expenditure (RD) and Resident Patents (ResPatent) did not originally include zeros, which reduced the sample size dramatically. Thus, we re-coded all missing data as zeros. This may have erroneously overlooked the R&D spending of some countries with negligible amounts. Notably, Taiwan lacked data on these measures, but we know it has sizeable Research and Development and Research Patent activity. But, lacking data we excluded it from the model. These choices resulted in a sample of 172 countries. Finally, among this sample, some variables had partial missing data for certain years but not others. In this situation, we imputed the mean of the available data for the country.

### 3.c. Methods of Analysis

Starting with the proposition that the chosen indicators are related to research capacity, we apply the EFA to identify their relationships. EFA is a statistical method used to identify unobserved "latent factors" that manifest numerous observable indicators (Cudeck, 2000). It is commonly used to justify the reduction of numerous variables into aggregate indexes. EFA computes the pairwise correlation matrix of a set of variables, then computes the eigenvalues and eigenvectors of the matrix, which are used to identify the amount of variance in the indicator variables explained by the factor (eigenvalues) and the direction of the relationship between the variables and the underlying factor (eigenvector). In the present context, we use EFA to identify whether the indicators listed in Table 1 (excluding FWCI) represent a coherent underlying latent factor, called "national research capacity". Practically, candidate variables are found in a variety of sources and formats, and they are often gathered for reasons unrelated to research capacity. As such, their relationship with one another becomes more important than what they represent individually. Since there are numerous variables that measure essentially the same factor, EFA helps to economize the multiplicity of empirical indicators. Factor analysis allows us to make statements about the convergence or divergence of these empirical measurements as they relate to national research capacity. We use the R package 'psych' to conduct EFA using the principal factor method and varimax rotation, retaining 2 factors as indicated by the eigenvalue and eigenvector matrixes (Grice, 2001).

The Cronbach's Alpha test, which examines the internal consistency and relatedness of a set of variables, is used to assess scale reliability (Revelle and Condon, 2019). This test is conducted after EFA to provide additional evidence that the variables identified by EFA have scale reliability prior to aggregation. Higher scores indicate greater internal consistency. In general, a value greater than 0.7 indicates adequate scale reliability. To aggregate variables into an index for the cross-sectional regression model, we chose the factor regression score extraction method. We used a summative index for the panel regression because there appears to be no apparent way to conduct factor analysis on panel data.

Additionally, we wish to demonstrate the relationship between the research capacity index and other well-established variables. To achieve this, we employ fractional Field Weighted Citation Impact (FWCI) to measure the impact of national research. FWCI is the ratio of the total citations received by the unit (country) and the total citations expected based on the average of the subject field (in this case all domains). The data are further fractionalized in cases of international collaboration representing country-specific contributions. FWCI has gained acceptance in the scientometrics literature as a valid indicator of citation impact and fractional counting is a growing standard for analysis (Waltmann and van Eck, 2015; Purkayastha et al., 2019; Sivertsen, Rousseau, & Zhang, 2019).

To examine the influence of the national research capacity index on FWCI, we employ Bayesian multilevel regression with the R package brms (Buerkner, 2017). This method allows us to account for the distinctions between regions and countries in our data (see also, Huggins & Izushi, 2008), revealing the relationship between research capacity and research impact across the globe. Bayesian methods have found rapid acceptance as an alternative to the frequentist approach of conventional regression techniques that employ significance testing based on the p-value. In place of a binary evaluation of statistical significance, Bayesian methods generate credibility intervals for parameter estimates of interest, focusing the analyst on gradations of uncertainty (Gelman, Hill, & Vehtari, 2020).

## 4. Results

So far, the objective has been to add governance factors to traditional R&D measures to expand measures of national research capacity for the purposes of constructing an index. This section summarizes the results of the statistical analysis. First, we present descriptive statistics. Second, we show the results of the exploratory factor analysis. Finally, we present the Bayesian regression models that predict research impact using the index.

Prior to presenting the findings, it is useful to provide some remarks on the empirical methodology employed. Longitudinal data were collected for all variables spanning the years 2013 to 2021. There are no methods for conducting exploratory factor analysis on longitudinal data. Consequently, we aggregated all the data based on the within-country mean for the time frame in order to conduct the analysis. Practitioners and analysts may choose different time frames or to construct indexes year-by-year as the data become available.

Table 2 displays the list of indicators along with descriptive statistics of the variables analyzed. The N shows the number of countries for which data are included. The mean, sd, min, and max present descriptive statistics for each indicator. The mean shows that the data have positive and negative values. The standard deviation (sd) shows how far the data points are from the mean. The min/max measures show the range of the data. Prior to conducting the exploratory factor analysis, the natural logarithm was applied to variables exhibiting significant skewness. All variables pertaining to raw research capacity, ranging from research and development (RD) to international publications (IntlPubs), were incorporated. This resulted in more normally distributed variables. The variables pertaining to governance exhibited less skewed distributions that did not require log transformation.

*Table 2 Descriptive Statistics of the Entire Data Set, 2013-2021*

| Variable | N | mean | sd | min | max |
|---|---|---|---|---|---|
| **ln_RD** | 172 | 14.455 | 9.228 | 0 | 27.068 |
| **ln_ResPatent** | 172 | 3.906 | 3.189 | 0 | 13.923 |
| **ln_AcadInst** | 172 | 3.063 | 1.586 | 0.105 | 7.677 |
| **ln_NonAcadInst** | 172 | 2.883 | 1.634 | 0.105 | 8.139 |
| **ln_Authors** | 172 | 7.792 | 2.478 | 2.485 | 14.002 |
| **ln_Pubs** | 172 | 6.736 | 2.774 | 0.896 | 13.183 |
| **ln_IntlPubs** | 172 | 5.813 | 2.373 | 0.849 | 11.437 |
| **OpenInternet** | 172 | 0.387 | 1.511 | -3.572 | 2.372 |
| **RuleLaw** | 172 | 0.555 | 0.303 | 0.017 | 0.998 |
| **RegQual** | 172 | -0.118 | 0.993 | -2.32 | 2.056 |
| **Stability** | 172 | -0.205 | 0.951 | -2.754 | 1.493 |
| **NonCorrupt** | 172 | -0.115 | 1.005 | -1.714 | 2.26 |
| **Polyarchy** | 172 | 0.527 | 0.25 | 0.017 | 0.919 |
| **AcadFreedom** | 172 | 0.638 | 0.289 | 0.025 | 0.971 |
| **FWCI** | 172 | 0.79 | 0.252 | 0.276 | 1.594 |

A correlogram for each variable is shown in Figure 1. The pairwise correlations for each pair of variables are displayed in the table's upper section. Each variable's distribution is represented by the diagonal. Each bivariate scatterplot is displayed in the table's lower section, along with fit lines that roughly depict the correlation's slope. The graph displays two regions of higher correlations, where 1) the governance measures and 2) the raw capacity metrics have stronger correlations with one another. The image also demonstrates the correlation between these two sets of metrics and FWCI. It appears that the governance measures have a stronger correlation with FWCI than the raw capacity measures, which will be discussed further below.

*Figure 1 Correlogram of all Variables*

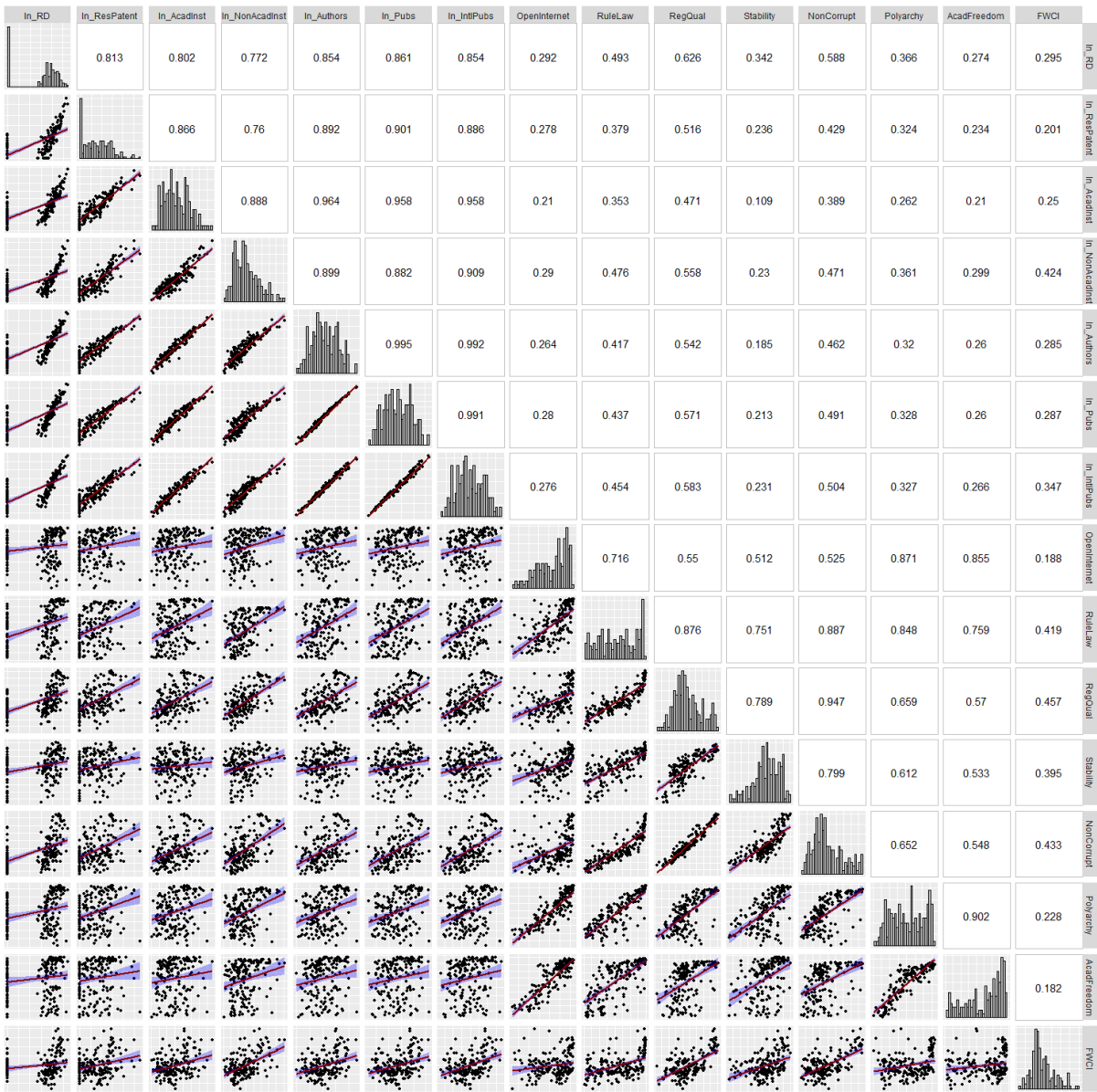

Two exploratory tests were performed to ensure that factor analysis would be a satisfactory tool to assess the relationship among the variables. The initial tests confirmed that the data are appropriate for factor analysis. First, the Kaiser-Meyer-Olkin factor adequacy test yielded a mean sampling of adequacy (MSA) value of 0.88. A value close to 1 suggests that the variable is highly correlated with other variables and is therefore suitable for factor analysis. The statistical analysis known as Bartlett's test of sphericity yielded a p-value of 0, indicating that the observed correlation matrix differs significantly from an identity matrix, further confirming that the variables under consideration are appropriate for factor analysis.

Following these results, the correlation matrix was tested with eigenvalue decomposition, revealing the presence of three factors with eigenvalues greater than 1. These factors possessed respective values of 8.11, 3.4, and 1.13. The scree plot in Figure 2 demonstrates that two factors are situated beyond the inflection point of the curve, while the third factor exhibits only a marginal increase above one. The scree plot displays the cumulative percentage of variance accounted for by each successive factor, indicating that two factors explain 82% of the variance in the variables. A third factor shows only a small increase in the percentage of variance explained above.

*Figure 2 Scree Plot*

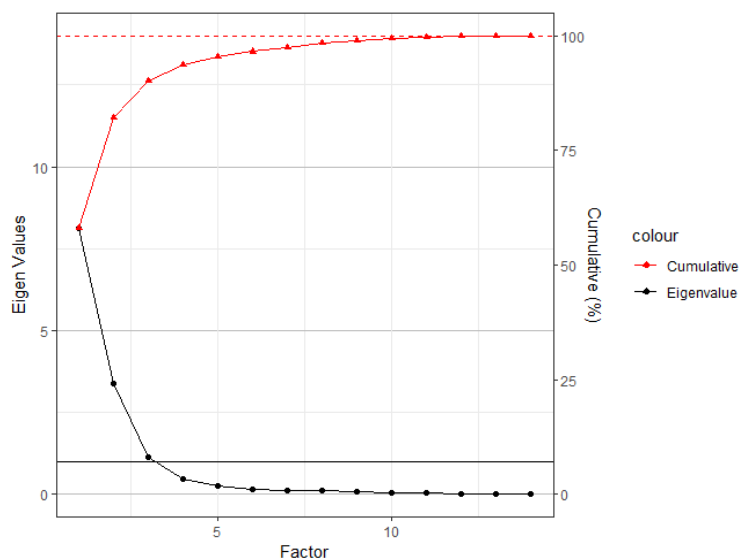

Following the initial testing, the EFA is demonstrated through the application of the 'fa' function in the 'psych' package within the R programming language. The model options selected are the principal factor method and varimax rotation. The standardized factor loadings presented in Table 3 illustrate the associations between the variables and the underlying factors in the first two identified columns. The factor loadings reveal the degree of correlation between each variable and the two factors identified by the EFA, showing the extent to which each variable is associated with or "loaded" onto the factors. The first factor's factor loadings for variables RD through IntlPubs exhibit a strong performance, as they all exceed 0.67. Factor two exhibits loadings that surpass 0.71 on the constructs of OpenInternet through AcadFreedom. The loadings indicate robust associations between the observed variables and the latent factors, and the pattern of loadings aligns with established theoretical frameworks.

Factor 1 appears to predict measures associated with core capability, whereas Factor 2 is associated with measures of governance. The term "h2" denotes the squared value of "h" and represents the shared variance that is explained by the respective variable. Stated differently, what proportion of the variability in the given variable can be attributed to the specific factor being considered? The notation "u2" denotes the square of the variable "u," and conveys its distinctiveness in relation to the inverse of "h2." Finally, the measure of complexity is determined by the ratio of h2 to u2, whereby larger values signify a greater degree of influence from shared factors on the variable in question. The RD metric stands out significantly in comparison to other core capacity measures, potentially attributable to the imputation of zero values in instances of missing data. Furthermore, it is noteworthy that bibliometric indicators exhibit a strong positive correlation, resulting in reduced explanatory power of the two econometric variables by the shared factor. The variables pertaining to governance exhibit a greater degree of heterogeneity, with Rule of Law, Regulatory Quality, and Polyarchy being the factors that are most strongly associated with the aforementioned variable.

Table 3 – Standardized Loadings, Communality, Uniqueness, and Complexity

| Factor | Factor 1 | Factor 2 | h2 | u2 | com |
|---|---|---|---|---|---|
| ln_RD | 0.668 | 0.213 | 0.491 | 0.509 | 0.965 |
| ln_ResPatent | 0.896 | 0.149 | 0.825 | 0.175 | 4.714 |
| ln_AcadInst | 0.962 | 0.089 | 0.933 | 0.067 | 13.925 |
| ln_NonAcadInst | 0.873 | 0.269 | 0.834 | 0.166 | 5.024 |
| ln_Authors | 0.981 | 0.153 | 0.987 | 0.013 | 75.923 |
| ln_Pubs | 0.978 | 0.166 | 0.985 | 0.015 | 65.667 |
| ln_IntlPubs | 0.974 | 0.195 | 0.987 | 0.013 | 75.923 |
| OpenInternet | 0.044 | 0.772 | 0.599 | 0.401 | 1.494 |
| RuleLaw | 0.274 | 0.917 | 0.916 | 0.084 | 10.905 |
| RegQual | 0.472 | 0.764 | 0.806 | 0.194 | 4.155 |
| Stability | 0.119 | 0.713 | 0.522 | 0.478 | 1.092 |
| NonCorrupt | 0.398 | 0.754 | 0.727 | 0.273 | 2.663 |
| Polyarchy | 0.146 | 0.911 | 0.852 | 0.148 | 5.757 |
| AcadFreedom | 0.009 | 0.814 | 0.663 | 0.337 | 1.967 |

From these results, we conclude that the first factor, composed of measures of national wealth expressed in research and development expenditure, resident patent registrations, numbers of authors, number of academic and non-academic institutions, number of publications, and international publications provide the most straightforward candidate measures for aggregation into an index representing national research capacity. The second factor appears to be a good candidate for aggregation into an index representing governance, or wider context, of the country as these variables relate to but do not directly contribute to capacity.

Next, we present the results of Cronbach's alpha test of scale reliability on the items loading separately on each factor. The text assesses the degree to which the items in a set are interrelated, testing whether they measure a single, underlying construct or dimension. Cronbach's Alpha ranges from 0 to 1, with higher values indicating higher average inter-item reliability. Typically, a Cronbach's Alpha value of 0.7 or higher is considered acceptable. The test results for the items loading on factor 1, representing raw capacity, including variables RD through IntlPubs, resulted in an overall alpha value of 0.85, a standardized alpha value of 0.98, and a Guttman's lambda 6 coefficient of 0.98. Similarly, items loading on factor 2, representing governance, including variables OpenInternet through AcadFreedom, resulted in an overall alpha value of 0.85, a standardized alpha value of 0.94, and a Guttman's value of 0.97. Tests for both sets of variables indicate moderate to high levels of internal reliability and suggest aggregation in indexes is appropriate.

To visualize how countries rank on raw capacity and governance, Figure 3 shows three plots. The left plot shows countries ranked from highest to lowest on core research capacity, using regression scores extracted from the factor analysis. The middle plot shows countries ranked on governance ("wider context") using factor regression scores. The right plot shows countries ranked by the interaction between (product of) these two scores. The scores were first standardized before plotting. Notably, the three plots present different rankings for lists of countries. Table 4 compares the top 10 country names in each plot in Figure 3.

*Table 4 Factor Score Country Ranks*

| Capacity Factor | Governance Factor | Capacity X Governance |
|---|---|---|
| China | Luxembourg | United States |
| United States | Iceland | Germany |
| India | Estonia | France |
| Russia | Denmark | Great Britain |
| Japan | Finland | Japan |
| Great Britain | Norway | Australia |
| Iran | New Zealand | Canada |
| Germany | Sweden | Switzerland |
| France | Switzerland | Netherlands |
| Brazil | Austria | Sweden |

*Figure 3 Factor Score Country Ranks*

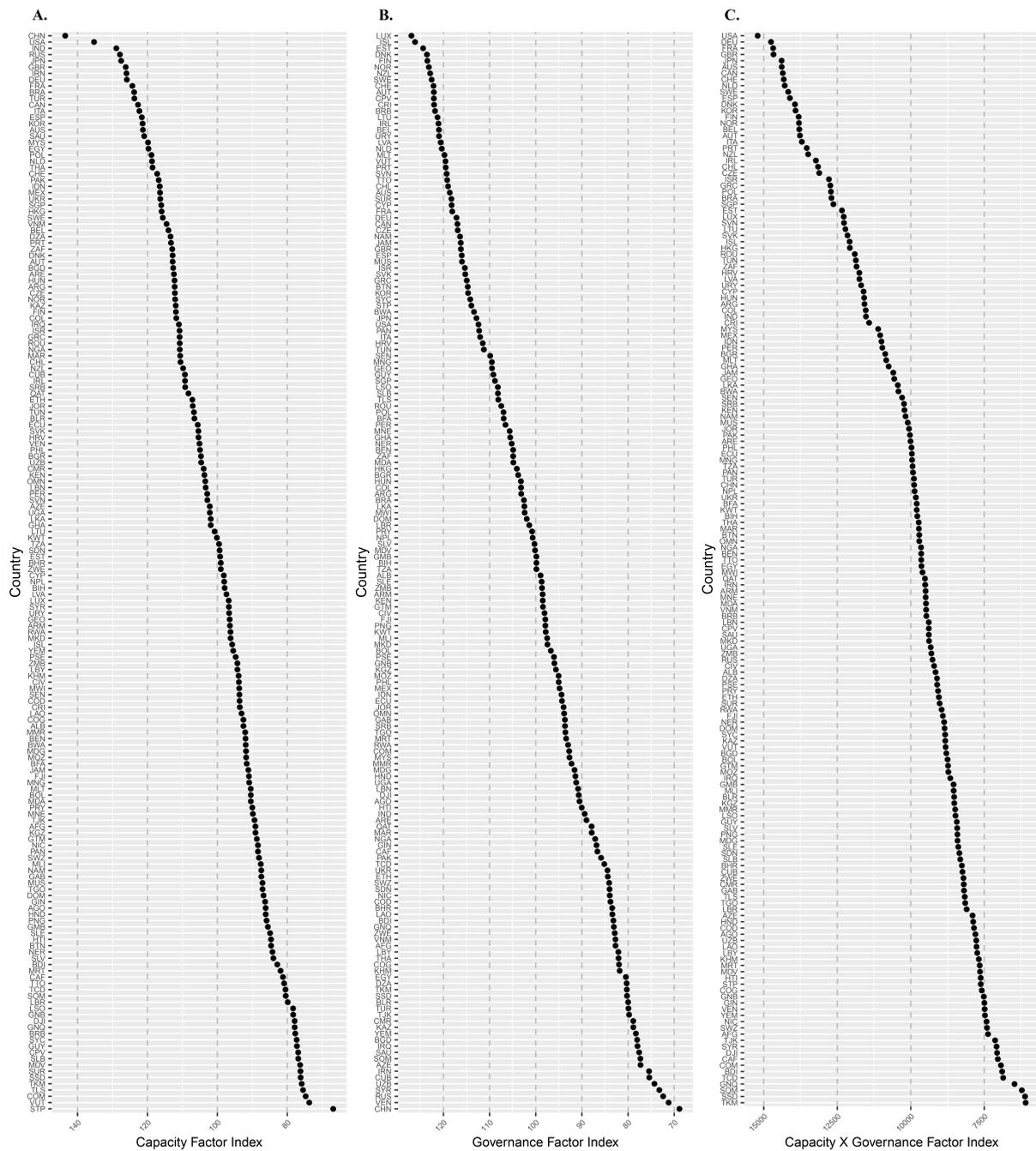

To illustrate how different aggregation methods result in different country rankings, the rankings are reconstructed using a simpler summative index. Some practitioners and scholars may find this approach more intuitive. Further, the results of the summative index are useful in the longitudinal data setting as will be demonstrated later in the paper.

*Table 5 Summative Index Country Ranks*

| Capacity Factor | Governance Factor | Capacity X Governance |
|-----------------|-------------------|------------------------|
| United States | Norway | United States |
| China | Sweden | Germany |
| Japan | Denmark | Great Britain |
| Germany | Finland | Japan |
| Great Britain | Switzerland | Canada |
| India | New Zealand | France |
| France | Luxembourg | Switzerland |
| South Korea | Iceland | Australia |
| Italy | Canada | Netherlands |
| Russia | Netherlands | Sweden |

*Figure 4 Summative Index Country Ranks*

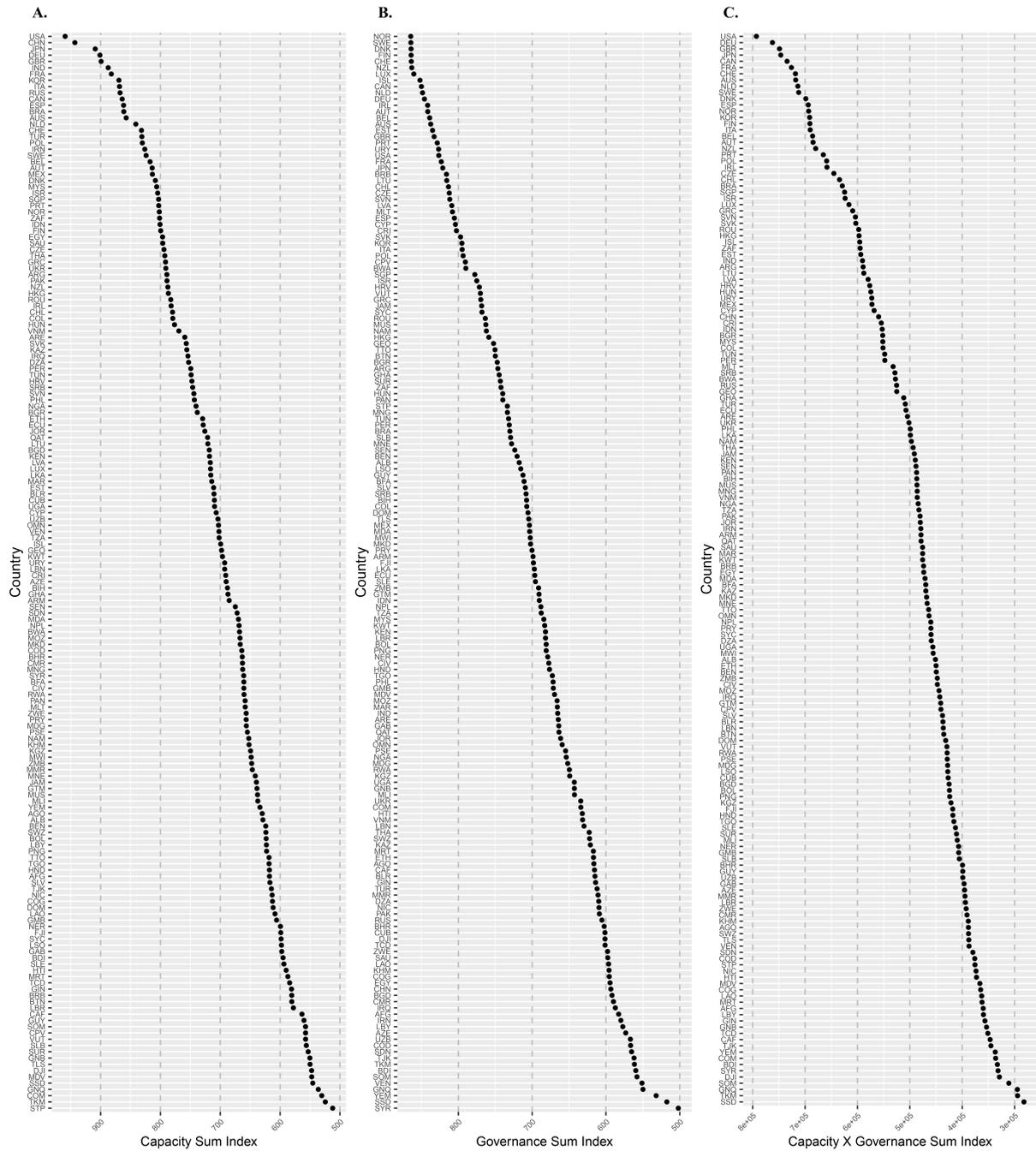

Finally, we move to test the usefulness of the two measures, drawn from the EFA, as they relate to a measure of national research output. Capacity and Governance, as predictors of national research impact, are regressed against the fractional FWCI. First, the results of a cross-sectional (averaged by country over 2013-2021) Bayesian mixed-model regression are shown in Table 6, which tests the effects of Capacity and Governance on country-level research impact measured via FWCI. To control for known issues with FWCI, it is common practice to filter out countries that have a very low publication rate but extremely high FWCI due to collaboration. Thus, we removed countries with less than 50 total publications, taking the model from a sample size of 172 to 157.

*Table 6 Cross-Sectional Bayesian Regression, Data Collapsed 2013-2021*

| FWCI | Estimate | Std. Error | [95% cred. interval] |
|---|---|---|---|
| Intercept | -0.0037 | 0.2166 | [-0.4315,0.4232] |
| Capacity | 0.0033 | 0.0014 | [0.0006, 0.0059] |
| Governance | 0.0047 | 0.0012 | [0.0023, 0.0072] |
| Region | 0.1753 | 0.0571 | [0.0950, 0.3153] |
| Residuals | 0.1762 | 0.0106 | [0.1573, 0.1982] |

Table 6 shows that both Capacity and Governance have positive estimates on fractional FWCI. Further, the standard errors are less than half the estimates and the credibility intervals and both do not include zero, indicating in frequentist terminology that the estimate is "statistically significant". The model also controls for 10 different geographic regions.

Table 7 shows the regression analysis conducted on the full panel data. Given the absence of a clear method for obtaining factor regression scores across temporal intervals, a summative composite measure was employed in lieu of extracting regression scores. The present model incorporates the hierarchical nature of the longitudinal dataset, encompassing data pertaining to individual countries over time, which are nested within regions. This facilitates the control of the influence of temporal and regional factors on the impact of research. The outcomes of the model exhibit a considerable resemblance to the findings obtained from the cross-sectional model. Table 7 displays affirmative evaluations on the dimensions of capacity and governance, with credibility intervals that exclude zero. Time is accounted for both in terms of the country- and region-specific random intercepts, as well as yearly fixed effects. The region, the country nested within the region, and the residual exhibit positive estimates and their credibility intervals do not encompass the value of zero.

*Table 7 Longitudinal Bayesian Regression, 2013-2021*

| FWCI | Estimate | Std. Error | [95% cred. interval] |
|---|---|---|---|
| Intercept | 0.4997 | 0.0911 | [0.3227, 0.6836] |
| Capacity | 0.0045 | 0.0013 | [0.0019, 0.0072] |
| Governance | 0.0177 | 0.0041 | [0.0097, 0.0257] |
| Time | Fixed | Fixed | Fixed |
| Region | 0.1598 | 0.0541 | [0.0876, 0.2957] |
| Region:Country | 0.1619 | 0.0118 | [0.1405, 0.1860] |
| Residuals | 0.1226 | 0.0028 | [0.1172, 0.1281] |

## 5. Discussion and Conclusion

The present study introduces a novel composite index of a country's capacity to produce research based on a set of credible indicators. The index is then tested to establish a correlation with national field-weighted citations. Two factors emerge from the data, one representing raw or "core" capacity and a second representing the governance or the "wider context" of the country's assets. The results show quite different rankings for countries based on core capacity versus governance context. China's position, for example, moves from first in core capacity, last of 158 countries in governance, and position 74 in the combined measure.

An index is an efficient way to combine indicators from various sources into a useful set. The index herein is designed to facilitate the policy process by combining existing data into valuable, comparable, and analogous collections. The included variables are drawn from the accessible databases. International business data has been deliberately omitted to improve feedback at the policy level. The index provides material for hypothesis testing about the roles of variables and governance in building the capacity of a country to produce useful knowledge.

Through various analyses and rigorous assessments, the article shows that a set of indicators are significantly associated with a country's research capacity. These indicators include research and development expenditures, resident patent registrations, author and publication counts, as well as international publications. These indicators have been aggregated into a composite index meeting the definition of Lundvall's "core capacity". The second factor that facilitates the support of the national research capacity study is operationalized as Lundvall's "wider context" or governance..

The results provide to policymakers and analysts the ability to compare nations against one another, and perhaps to consider asymmetries between countries.. Moreover, actions within the sphere of "science diplomacy" may be helped by this approach when actions involve establishing scientific agreements or proposing ties. Policymakers sometimes lack clear insight into the underlying capacities of counterpart nations as they seek partners to participate in scientific activities, and this index may be of help since soliciting science agreements or proposing ties can at times be opaque, particularly with regard to the least developed nations.

This index differs from the GII, the GKI, and the GCI by avoiding multilevel variables and nested indices, which reduce transparency. The research capacity index draws upon known data from solid sources, and it includes analysis over eight years. Further, by removing globalized and economic variables, the proposed index provides more accurate feedback to policymakers and evaluators who are concerned about the health of national research capacity and its application.

The index may also be useful for countries wishing to promote their scientific investments and achievements. The analysis tested and showed strong relationships for the raw or core indicators as those most related to capacity. National capacity to conduct research and development can attract talent who wish to cooperate or collaborate, invest, or study in another country. Nations with higher research capacity attract students and researchers to their universities and research institutions. Governments approve investment into R&D to build research capacity to reach multiple goals, which may include systemic resilience, long-term viability, and national standing and prestige in science. Understanding the role of core capacity and wider context may aid policymakers as they consider ways to improve the development of "useful knowledge."

The capacity index could certainly be improved in the future. The focus here is on the underlying latent factors that manifest as relationship between indicators and less about the specific indicators that go into an index. In social science underlying causal mechanisms generate innumerable empirical indicators, and we must not lose sight of the forest for the trees. In this case, the forest represents the more or less stable relationship between indicators, while the indicators themselves represent the individual trees.

Additional research could seek to validate the index and assess the scope for trimming or expanding additional indicators such as tertiary education levels, tax incentives, infrastructure, and an improved citation measure. The index could become more useful over time as more data points are added. In addition, further research using the index in inferential models on a wide variety of interesting outcomes, such as strategic behavior in the international system, may provide insights into the effects of the capacity of individual nations on their network of relationships.

We hope that the index acts as a useful tool for assessing current science capacity and encouraging international collaboration. But more importantly, we hope our approach may serve as a practical starting point for other scholars seeking to construct their own indexes. We expect to use it for research to understand the influence of geopolitical factors on national growth and international collaboration. Further, we expect to use the index to serve as a test for the role of public investment in the growth of capacity over time.

## Acknowledgments

Special thanks to Edwin Horlings for comments on the analytical and conceptual approach. Thanks to Ian Helfrich and Lisa Fagan for comments on the statistical analysis. Thanks to Jeroen Baas at Elsevier for providing critical data. Thanks also to Carol Robbins, Sylvia Schwaab-Serger, and John Jankowski, and the late Loet Leydesdorff for consultations and comments; and to Peter Zhang and Ken Poland for comments on data collection. An earlier monograph by Wagner et al. (2001) presents a similar approach to indexing national research capacity, so we acknowledge the previous RAND publication. We are thankful to attendees at the Atlanta Conference on Science and Innovation Policy, 2023, for comments.